\documentstyle[preprint,aps]{revtex}
\begin{document}
\draft

\title{Hamiltonian in Financial Markets}

\author{Jun-ichi Maskawa}
\address{Department of Management Information, Fukuyama Heisei University, Fukuyama, Hiroshima 720-0001, Japan}
\maketitle
\tightenlines

\begin{abstract}

A statistical physics model for the time evolutions of stock portfolios is proposed. 
In this model the time series of price changes are coded into the sequences of up and down spins.
The Hamiltonian of the system is introduced and is expressed by spin-spin interactions as in spin glass models of disordered magnetic systems. 
The interaction coefficients between two stocks are determined by empirical data coded into up and down spin sequences using fluctuation-response theorem.
Monte Carlo simulations are performed and the resultant probability densities of the system energy and magnetization show good agreement with empirical data.

\end{abstract}

The data analysis and modeling of financial markets have been hot research subjects for physicists as well as economists and mathematicians in recent years\cite{econo}.
The non-Gaussian property of the probability distributions of price changes, in stock markets and foreign exchange markets, has been one of main problems in this field\cite{econo}\cite{mandel}\cite{dist}. 
From the analysis of the high-frequency time series of market indices, e.g., S\&P500, Nikkei225, a universal property was found in the probability distributions.
The central part of the distribution agrees well with Levy stable distribution\cite{levy}, while the tail deviate from it and shows another power law asymptotic behavior.
The scaling property on the sampling time interval of data is also well described by the crossover of the two distributions.  
Several stochastic models of the fluctuation dynamics of stock prices are proposed, which reproduce power law behavior of the probability density\cite{econo}\cite{mandel}\cite{model}. 
The auto-correlation of financial time series is also an important problem for markets. 
There is no time correlation of price changes in daily scale, while from more detailed data analysis an exponential decay with a characteristic time $\tau=4$ minutes was found\cite{econo}\cite{dist}. 
The fact that there is no auto-correlation in daily scale is not equal to the independence of the time series in the scale.
In fact there is auto-correlation of volatility (absolute value of price change) with a power law tail\cite{econo}\cite{vola}.
Recently, the cross-correlation between pairs of stock issues was deeply investigated for the time series of price changes, 
and the hierarchical structure in the subdominant ultrametric space\cite{ultra} was found\cite{econo}\cite{cross}. 
This result suggests the complex collective time evolution of financial markets as in frustrated disordered systems like spin glass\cite{spin} in which the ultrametricity has been established.
Those problems listed here have been studied using the methods and the concepts developed in material sciences especially in the studies of complex systems.

In this paper, a statistical physics model for the collective time evolutions of stock portfolios is proposed. Portfolio is a set of stock issues. 
In this model we deal with the time series of price changes coded into the sequences of up and down spins.
A sample of coding procedure is shown in Fig. \ref{fig1}.
The Hamiltonian of the system is introduced and is expressed by spin-spin interactions as in spin glass models of disordered magnetic systems. 
The interaction coefficients between two stocks are phenomenologically determined by empirical data.
They are derived from the covariance of sequences of up and down spins using fluctuation-response theorem.
We investigate the stocks listed in Dow-Jones industrial average as a portfolio for the test of this model. 
Monte Carlo simulations using Gibbs weight as a transition probability reproduce the probability densities of the energy and the magnetization of the portfolio,
whose definitions are given later. 

We start with the Hamiltonian expression of our system that contain N stock issues. 
It is a function of the configuration $S$ consisting of N coded price changes $S_i \enspace (i=1,2,...,N)$ at equal trading time.
The interaction coefficients are also dynamical variables, because the interactions between stocks are thought to change from time to time.
We divide a coefficient into two parts, the constant part $J_{ij}$, which will be phenomenologically determined later, and the dynamical part $\delta J_{ij}$.
The Hamiltonian including the interaction with external fields $h_i \enspace (i=1,2,...,N)$ is defined as    
\begin{equation}
H[S,\delta J,h]= \sum_{<i,j>} [\frac{ \delta J_{ij}^2}{2 \Delta_{ij}} - (J_{ij} + \delta J_{ij})S_i S_j)]-\sum_i h_i S_i.
\label{h}
\end{equation}
The summation is taken over all pairs of stock issues.
This form of Hamiltonian is that of annealed spin glass\cite{spin}. 
The fluctuations $\delta J_{ij}$ are assumed to distribute according to Gaussian function. 
A statement referring to the justification of this assumption will be made later (see Fig. \ref{fig6}). 
The main part of statistical physics is the evaluation of partition function that is given by the following functional in this case  
\begin{equation}
Z[h]=\sum_{\{S_i\}}\int \prod_{<i,j>}\frac{d \delta J_{ij}}{(2 \pi \Delta_{ij})^{1/2}} \exp (-H[S,\delta J,h]).
\end{equation}    
The integration over the variables $\delta J_{ij}$ is easily performed and gives
\begin{equation}
Z[h]=A\sum_{\{S_i\}} \exp (-H_{eff}[S,h]).
\label{zav}
\end{equation}    
Here the effective Hamiltonian $H_{eff}[S,h]$ is defined as
\begin{equation}
H_{eff}[S,h]=-\sum_{<i,j>}J_{ij}S_i S_j-\sum_i h_i S_i,
\label{heff}
\end{equation}    
and $A=\exp (1/2\sum\Delta_{ij})$ is just a normalization factor which is irrelevant to the following step.
This form of Hamiltonian with constant $J_{ij}$ is that of quenched spin glass\cite{spin}. 

The constant interaction coefficients $J_{ij}$ are still undetermined.
We use fluctuation-response theorem which relates the susceptibility $\chi_{ij}$ with the covariance $C_{ij}$ between dynamical variables in order to determine those constants, which is given by the equation 
\begin{equation}
\chi_{ij}=\left.\frac{\partial m_i}{\partial h_j}\right|_{h=0}=C_{ij}.
\label{chi}
\end{equation}    
The notations in this paper are $m_i=<S_i>$ and $C_{ij}=<S_i-m_i><S_j-m_j>$.
For the evaluation of the left side of the equation (\ref{chi}), we use Thouless-Anderson-Palmer (TAP) equation for quenched spin glass\cite{tap}
\begin{equation}
m_i=\tanh (\sum_{j}J_{ij}m_j + h_i - \sum_{j}J_{ij}^2(1-m^2_j)m_i).
\label{mi}
\end{equation}    
We will neglect the third term of self-reaction \cite{reaction} and the nonlinear terms $\sim O(m^2_i)$\cite{hi} in the equations (\ref{mi}).
The equation (\ref{chi}) and the linear approximation of the equation (\ref{mi}) yield the equation 
\begin{equation}
\sum_{k}(\delta_{ik}-J_{ik})C_{kj}=\delta_{ij}.
\label{jik}
\end{equation}
Interpreting $C_{ij}$ as the time average of empirical data over a observation time rather than ensemble average,
the constant interaction coefficients $J_{ij}$ is phenomenologically determined by the equation (\ref{jik}).

We investigate two data sets of a portfolio containing $N=30$ stock issues listed in Dow-Jones industrial average for the test of this model.
One is the time series in the period from 16-May-2000 to 21-Jun-2000 and the other is in the period from 15-Aug-2000 to 26-Oct-2000.
The number of data amounts to about $30,000\times30=900,000$. 
The sampling interval is 1-minute, i.e., the interval two successive time stamps of samples are 1-minute. 
The time series of price changes are coded into the sequences of up and down spins (see Fig. \ref{fig1} again).
This coding is thought to be a kind of coarsening. The covariance $C_{ij}$  for 435 pairs of i and j are derived from the coded data.
Then the interaction coefficients $J_{ij}$ are calculated by the equation (\ref{jik}).
The energy spectra of the system, simply the portfolio energy, is defined as the eigenvalues of the Hamiltonian $H_{eff}[S,0]$.
The probability density of the portfolio energy can be obtained in two ways.
We can calculate the probability density from data by the equation 
\begin{equation}
p(E)\Delta E=P(E-\frac{\Delta E}{2} \le E \le E+\frac{\Delta E}{2}).
\label{pe}
\end{equation}
The results for two data sets are shown in Fig. \ref{fig2} ($\bigcirc$) with the results of Monte Carlo simulations ($\Box$) that will be explained below.   
The use of the theoretical equation
\begin{equation}
p(E)\Delta E=n(E)\frac{e^{ -E}}{Z[0]}
\label{gf}
\end{equation}
is the other way.
The function $n(E)$ is the density of states, which can be numerically obtained by random sampling from $2^{30}$ configurations.
The comparison between the two results is given in Fig. \ref{fig3}.
The theoretical lines explain well the empirical data except the ranges $E<-4$ and $1<E$ of small numbers of events,
indicating the canonical distribution of the system configurations S as $P(S)=\exp(-H_{eff}[S,0])/Z[0]$.

For another test of our model, Monte Carlo simulations using Gibbs weight as a transition probability are performed.
The transition probability from a configuration $S$ to $S'$ is given as
\begin{equation}
w(S\rightarrow S') = \left\{
	\begin{array}{cc}
		\exp(-\delta H) & if \enspace \delta H >0 \\
		& \\
		1 & otherwise,
	\end{array}
	\right.
\end{equation}
where $\delta H = H_{eff}[S',0]-H_{eff}[S,0]$. 
Monte Carlo simulations for the probability density of the portfolio energy and the system magnetization $m=(1/N)\sum_{i=1}^{N}S_i$ are given in Fig. \ref{fig2} ($\Box$) and Fig. \ref{fig4} ($\Box$) respectively.
They also show good agreement with empirical data. 
For the final consistency check of our model, we calculate the covariance $C^{mc}_{ij}$ between pairs of two stocks by Monte Carlo simulation. 
The frequency distribution of $\delta=|(C_{ij} - C^{mc}_{ij})/C_{ij}|$ is given in Fig. \ref{fig5}, which indicate the reproduction of $C_{ij}$ to the extent of our approximation.

Here I want to refer to the justification of the first term in the equation (\ref{h}), 
which indicates that the fluctuations $\delta J_{ij}$ distribute according to Gaussian function.
The direct verification of this assumption is impossible, because $\delta J_{ij}$ change from time to time.
Instead if we take the observation time long enough, the justification of this assumption is mathematically obtained by central limit theorem.
For the purpose of the test whether our observation time is long enough, we put the following test to the data in the period from 15-Aug-2000 to 26-Oct-2000.
First we take successive 50 trading days from the data, and then divide it into 10 pieces.
Next we derive $J^k_{ij}$ ( $k=1,...,10$) for each piece of data to obtain the distribution of $435\times10$ $\delta J^k_{ij}=J^k_{ij}-J_{ij}$.
The result is shown in Fig. \ref{fig6} supporting the statement of central limit theorem.      

In this paper, we gave a fully consistent phenomenological model for stock portfolios, which is expressed by the effective Hamiltonian (\ref{heff}).
This model will be also applicable to other financial markets that show collective time evolutions, 
e.g., foreign exchange market, options markets, inter-market interactions.
In our model, however, $J_{ij}$ must be determined by data via the calculation of $C_{ij}$.
The problem how we can go beyond this phenomenological standing point is unsolved.
There may be a hint in the universal property of the covariance matrix ${\bf C}=[C_{ij}]$ \cite{rmt} or (and) the existence of the correlation between a pair of $N(N-1)/2$ $J_{ij}$ corresponding to two separated time periods investigated above\cite{distofj}.


\newpage

\begin{figure}
\caption{A sample of the coding procedures for a virtual data. The first column is the serial number of data sampling, in which 0 means the time origin of observation. 
Samplings go along trading time at a fixed interval, e.g., sampling every 1-minute. From the price of i-th sample $Y(i)$, the i-th price change is defined as $Z(i)=Y(i)-Y(i-1)$.
The i-th code $S(i)=1$ (if $Z(i)>0$), $S(i)=-1$ (if $Z(i)<0$) and $S(i)=S(i-1)$ (if $Z(i)=0$).}
\label{fig1}
\end{figure}

\begin{figure}
\caption{The probability density $p(E)$ of portfolio energy $E$. 
$\bigcirc$: The empirical probability density ($\Delta E=0.1$ in the equation (\ref{pe})).
$\Box$: Monte Calro simulation of $2^{17}$ samples, which is explained later. 
{\bf (a)} The result for $9994\times30$ coded data of the time series sampled at 1-minute time interval in the period from 16-May-2000 to 21-Jun-2000.
{\bf (b)} The result for $19992\times30$ coded data of the time series sampled at 1-minute time interval in the period from 15-Aug-2000 to 26-Oct-2000.}
\label{fig2}
\end{figure}

\begin{figure}
\caption{Semi-log plot of the probability weight $p(E)n(0)/p(0)n(E)$ of portfolio energy $E$. 
The probability density $p(E)$ is the empirical result shown in Fig. 2, and the density of states $n(E)$ is numerically obtained by $2^{17}$ random sampling from $2^{30}$ configurations.
$\bigcirc$: The empirical probability weight.
{\it Solid line}: Gibbs weight $e^{-E}$ (no fitting parameter).
{\bf (a)} The result for the same data as in Fig. 2(a).
{\bf (b)} The result for the same data as in Fig. 2(b).}
\label{fig3}
\end{figure}

\begin{figure}
\caption{The probability density of the system magnetization defined in text for the coded data in the period from 15-Aug-2000 to 26-Oct-2000.
$\bigcirc$ :The probability density whose mean $\mu=-0.0134$, variance $\sigma^2=0.0744$.
{\it Dashed line}: Gaussian function with the same mean and variance as of the probability density.
$\Box$: Monte Calro simulation of $2^{19}$ samples}
\label{fig4}
\end{figure}

\begin{figure}
\caption{The frequency distribution of $\delta$ defined in text.}
\label{fig5}
\end{figure}

\begin{figure}
\caption{The probability density of the distribution of $435\times10$ $\delta J^k_{ij}$ whose definition is given in text.
$\bigcirc$: The probability density whose mean $\mu=-0.000324$, variance $\sigma^2=0.000679$, skewness $\beta_1=-0.0467$ and kurtosis $\beta_2=0.0604$.
{\it Dashed line}: Gaussian function with the same mean and variance as of the probability density.}
\label{fig6}
\end{figure}


\begin{references}
\bibitem{reaction}
In our system, the self-reaction term is negligible, because $J_{ij} \sim O(N^{-1})$\cite{distofj}.
I know this discussion is subtle.
The results of Monte Calro simulations that will be given later, however, seem to support this approximation.
\bibitem{hi}
In high-temperature phase, the linear approximation of the equation (\ref{mi}) is applicable.
It is plausible that our system mentioned here is in high-temperature phase, because stock prices would continue increasing or decreasing if it were in frozen phase.
The nonlinear terms may become crucial for the markets just before crash.
\end{references}
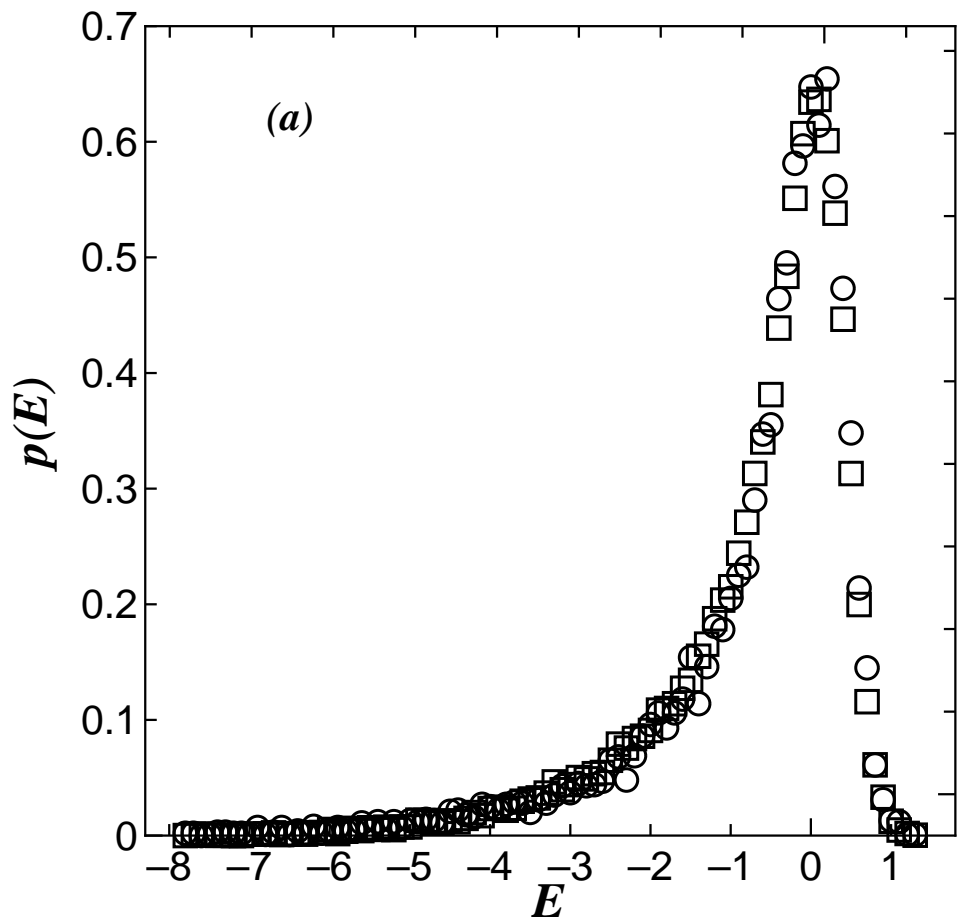
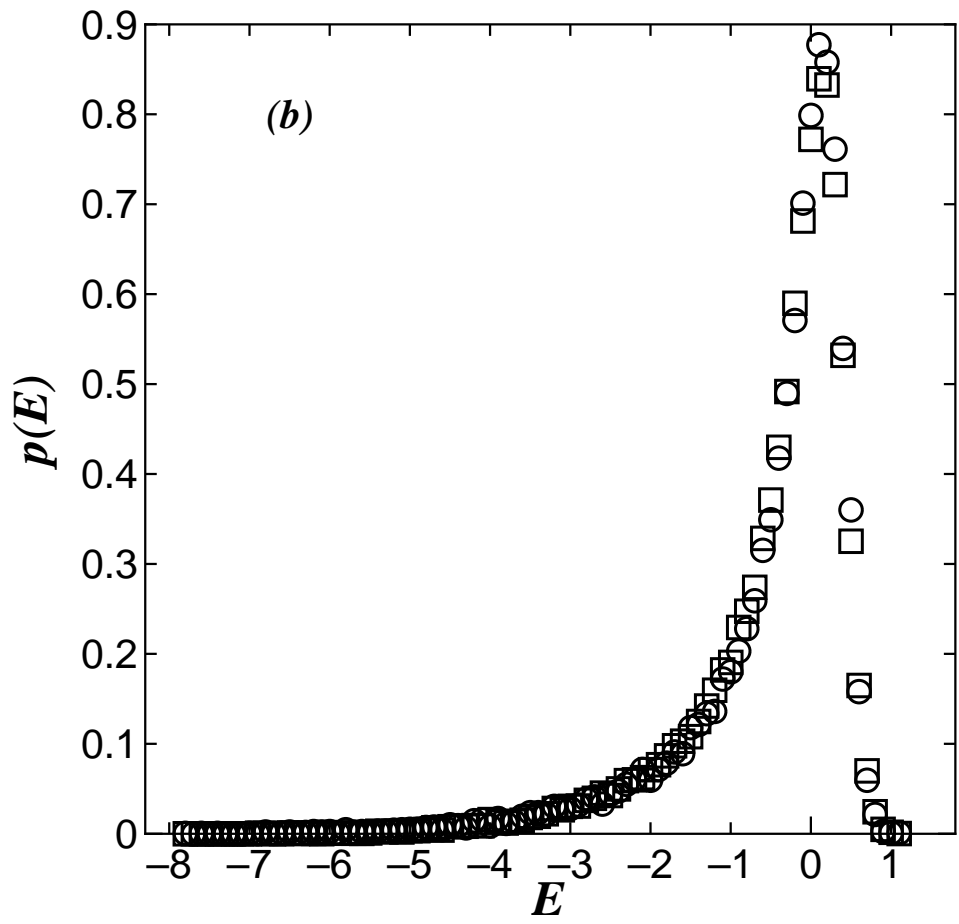
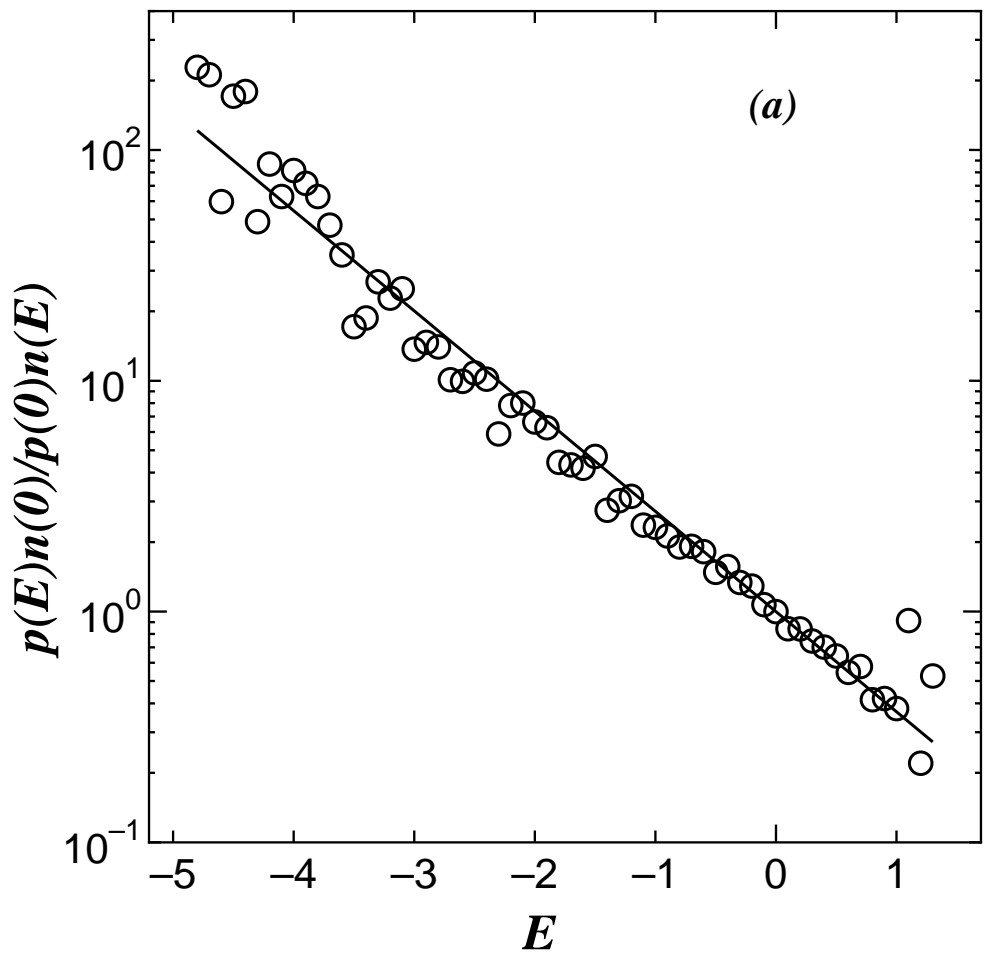
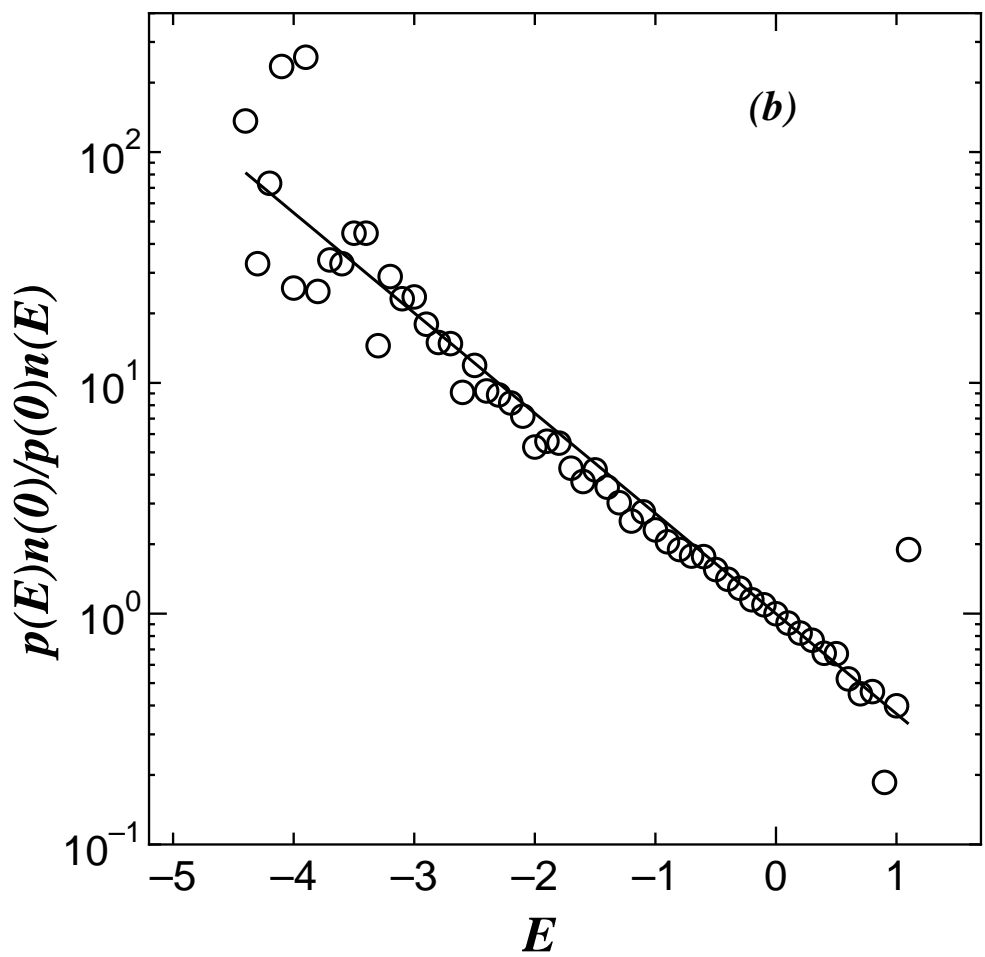
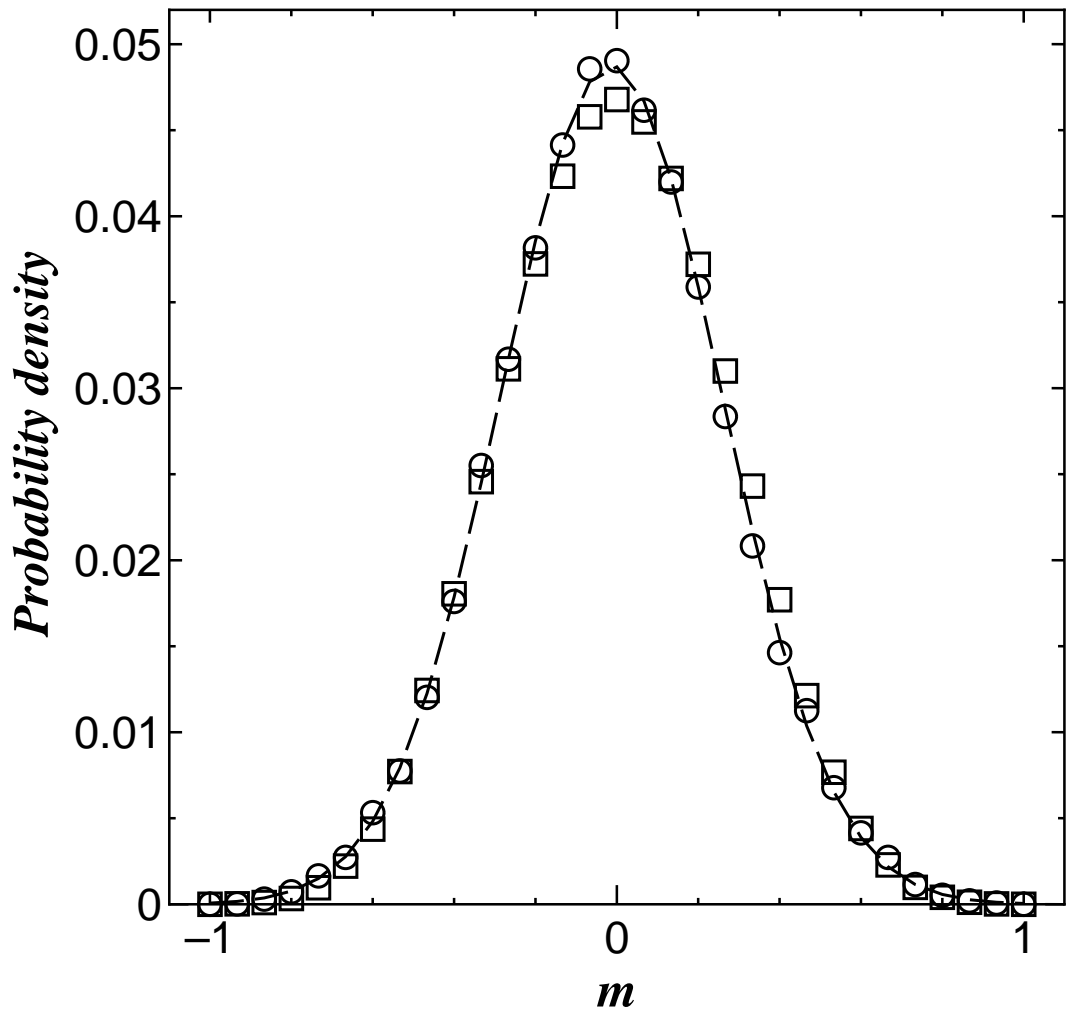
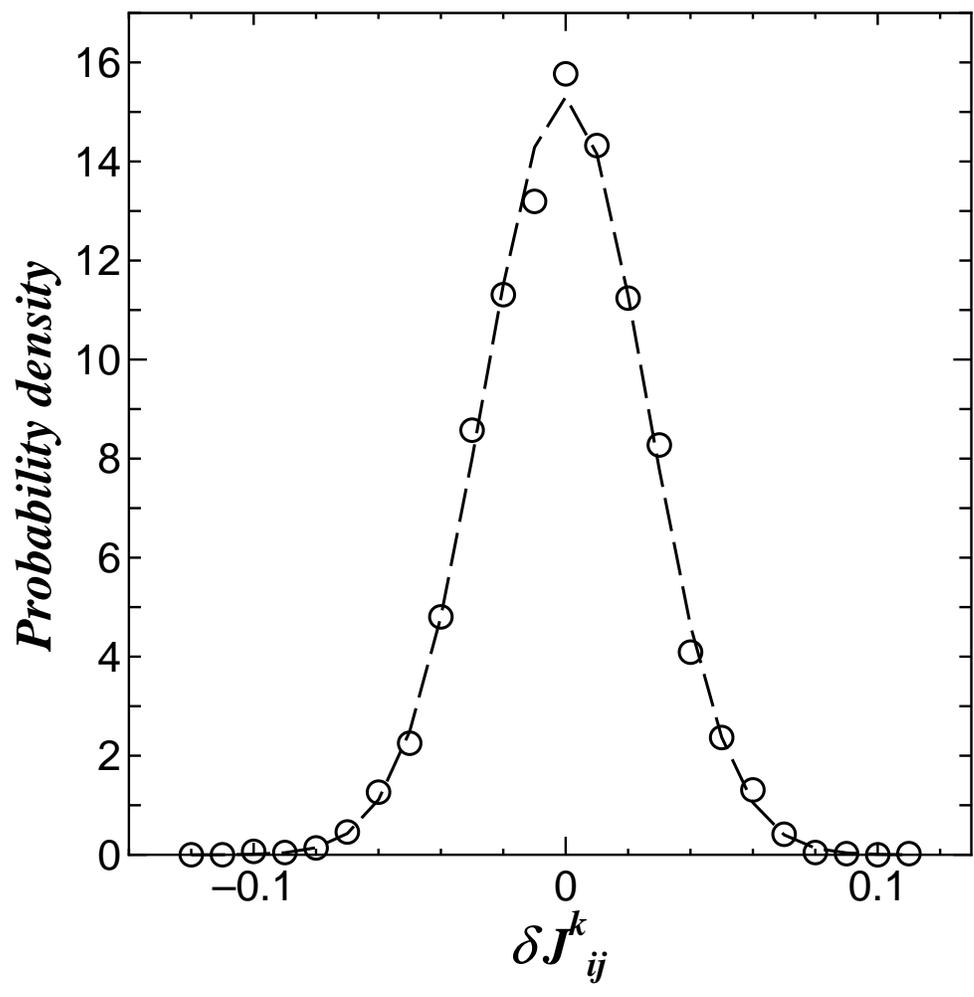
\end{document}